\documentclass[%
 reprint,
amsmath,amssymb,
aps,showpacs,
]{revtex4-1}
\usepackage{longtable}
\usepackage{graphicx}% Include figure files
\usepackage{bm}% bold math
\begin{document}

\preprint{APS/123-QED}

\title{Elasticity of a semiflexible filament with a discontinuous
  tension due to a cross-link or a molecular
  motor}% Force line breaks with \\

\author{Mohammadhosein Razbin$^{1}$}
\email{m.razbin@theorie.physik.uni-goettingen.de}
\author{Panayotis Benetatos$^{2}$}
\email{pben@knu.ac.kr}
\author{Annette Zippelius$^{1}$}

\affiliation{%
 $^{1}$ Institute for
  Theoretical Physics, Georg August University, Friedrich-Hund-Platz 1, 37077 G{\"o}ttingen, Germany\\
  $^{2}$ Department of Physics, Kyungpook National University, 80 Daehakro, Bukgu, Daegu 702-701, Republic of Korea
}%

\date{Published in Physical Review E (18 May 2016)}% It is always \today, today,
\begin{abstract}
We analyze the stretching elasticity of a wormlike chain with a tension discontinuity resulting from a Hookean spring connecting its backbone to a fixed point. The elasticity of isolated semiflexible filaments has been the subject in a significant body of literature, primarily because of its relevance to the mechanics of biological matter. In real systems, however, these filaments are usually part of supramolecular structures involving cross-linkers or molecular motors which cause tension discontinuities. Our model is intended as a minimal structural element incorporating such a discontinuity. We obtain analytical results in the weakly bending limit of the filament, concerning its force-extension relation and the response of the two parts in which the filament is divided by the spring. For a small tension discontinuity, the linear response of the filament extension to this discontinuity strongly depends on the external tension. For large external tension $f$, the spring force contributes a subdominant correction $\sim 1/f^{3/2}$ to the well known $\sim 1/\sqrt{f}$ dependence of the end-to-end extension.

\end{abstract}

%\keywords{Published in physical Review E (18 May 2016)}%Use showkeys class option if keyword
\maketitle

%\tableofcontents

\section{Introduction}

Over the past couple of decades, it has become clear that mechanics plays a very important role in the biological function of the cell, on a par with biochemistry \cite{Discher_Cell2006,Butcher_cancer_Nature2009,Ingber_Mechanics_Development2010}. In order to unravel the physical basis of the complex mechanical behaviour of biological matter, a bottom-up strategy has proven very fruitful \cite{bausch2006bottom}. In this approach, the basic functional modules of the cytoskeleton are reconstituted {\it in vitro} and analysed experimentally in tandem with theoretical modelling. 

The basic structural elements of the cytoskeleton (microtubules, intermediate filaments, F-actin) are all semiflexible polymers with a behaviour intermediate between that of a random coil and a rigid rod \cite{Terentjev_SM_review,Broedersz_MacKintosh_RMP}. They form supramolecular assemblies ({\it e.g.} networks, bundles) through cross-linking \cite{Bausch_Claessens_SM_review}. Cross-linking involves a host of different filament-binding proteins \cite{Ayscough_ABPs,Mandelkow_MAPs}. Active processes in the cell, such as the delivery of cargos, transport of organelles, mitotic dynamics, as well as muscle contraction are carried out by molecular motors using actin filaments or microtubules as tracks. The bottom-up approach to the study of molecular motors aims at analysing the transduction of metabolic energy into mechanical force and motion at the microscopic level using  {\it in vitro} assays \cite{Veigel_Schmidt_NR2011}. The advances in single-molecule manipulation are harnessed to study the simplest motor-filament complexes. In gliding assays, the motor (myosin, kinesin, or dynein) is attached to a glass surface and the translocation of the filament (F-actin or microtubule) is observed. In single motor assays, the filament is attached to the glass surface and the movement of the motor is monitored. In motor assays with beads, the motor is attached to a micron-sized refractile bead whose position is measured \cite{Holzbaur_Goldman_2010}. 

In this paper, we investigate analytically the mechanical response of
a semiflexible filament with a tension discontinuity. Our model system
can be viewed as one of the simplest structural elements of the
cytoskeleton beyond the isolated single-molecule level. We consider a
semiflexible polymer, modelled as a wormlike chain, in the weakly
bending approximation. The latter can be satisfied either by applying
a strong tensile force which irons out large thermal undulations, or
by having a filament with large persistence length compared to its
contour length. A longitudinal Hookean spring whose one end is
attached to a fixed substrate, has its other end on the filament thus
exerting a force which causes the tension discontinuity. The position
of one end of the filament is held fixed, whereas that of the other
end fluctuates. Its average position yields the force-extension
relations which are the main subject of our analysis. The spring may
be viewed as representing a motor, according to the myosin
cross-bridge model first introduced by Huxley in 1957
\cite{Huxley_1957} and still in use
\cite{Guerin_Prost_Martin_Joanny}. Our results apply to passive motors
or to very slowly stepping motors, slower than the relaxation time of
the filament. The time scale can be tuned by adjusting the
concentration of ATP molecules. We should point out, however, that our
study of semiflexible filaments with tension discontinuity is also
relevant to passive cross-linkers of large size or compliance. Many of
the actin binding proteins fall in this category as they can have
large spacer domains \cite{Bausch_Claessens_SM_review}

Our paper is organised as follows. In Sec. II we introduce the model and show how the discontinuity along the filament contour arises. In Sec. III we discuss the method of Green functions which allows us to calculate correlation functions of the filament tangent vector and therefore the force-extension relation for different boundary conditions. Explicit results for the case of small spring force, large spring force, and large spring and external forces with a small difference between them are given in Sec. IV. In Sec. V we discuss the relation of our analysis to single motor experiments. We conclude in Sec. VI, and details of our calculations are given in the Appendices.

\section{Model description}

 Our starting point is the wormlike chain model for a semiflexible
 filament of contour length $L$ which is attached to a point or wall on one
 side and pulled by a constant force, $\mathbf{f_{ext}}$, on the other
 side. Its Hamiltonian is given by
\begin{equation}
H_{WLC}=\frac{\kappa}{2}\int_{0}^{L}
(\frac{d\mathbf{t}}{ds})^2-\mathbf{f_{ext}}\int_{0}^{L}\frac{d\mathbf{t}}{ds}
\end{equation}
Here $\mathbf{t}(s)$ is the tangent vector at arclength $s, 0\leq
s\leq L$ and $\kappa$ denotes the bending stiffness. We will treat the
above model in the weakly bending approximation \cite{Marko_Siggia}, assuming that the
persistence length $l_p=\kappa/(k_B T)$ is much larger than the length
$L$ of the filament.

In the Monge parametrization~\cite{PhysRevE.74.041803}, the tangent of a semiflexible filament
is given by $t(s)=\frac{1}{\sqrt{1+a_1^2(s)+a_2^2(s)}}(1,a_1(s),a_2(s))$. We assume that
the filament is oriented along the x-direction, either due to the
grafting on the left side and/or the pulling force
$\mathbf{f_{ext}}=f_{ext}\mathbf{e}_x$.  In the weakly bending limit the
transverse components of the tangent vector $a_1(s)$ and $a_2(s)$ are small. We therefore approximate
$t_x(s)=1-\frac{1}{2}[a_1^2(s)+a_2^2(s)]$ and
$(\frac{dt(s)}{ds})^2=\dot{a}^2_1(s)+\dot{a}^2_2(s)$ where the dot
denotes the derivative with respect to $s$. The Hamiltonan then
reads
 \begin{equation}
H_{WB}=\sum_{i=1}^{2}\left[\int_{0}^{L}\left(\frac{\kappa}{2}\dot{a}^2_i(s)
+\frac{f_{ext}}{2}a_i^2(s)\right) ds\right]
\end{equation}

  \begin{figure}[th]
    \centering{
\includegraphics[width=0.50\textwidth]{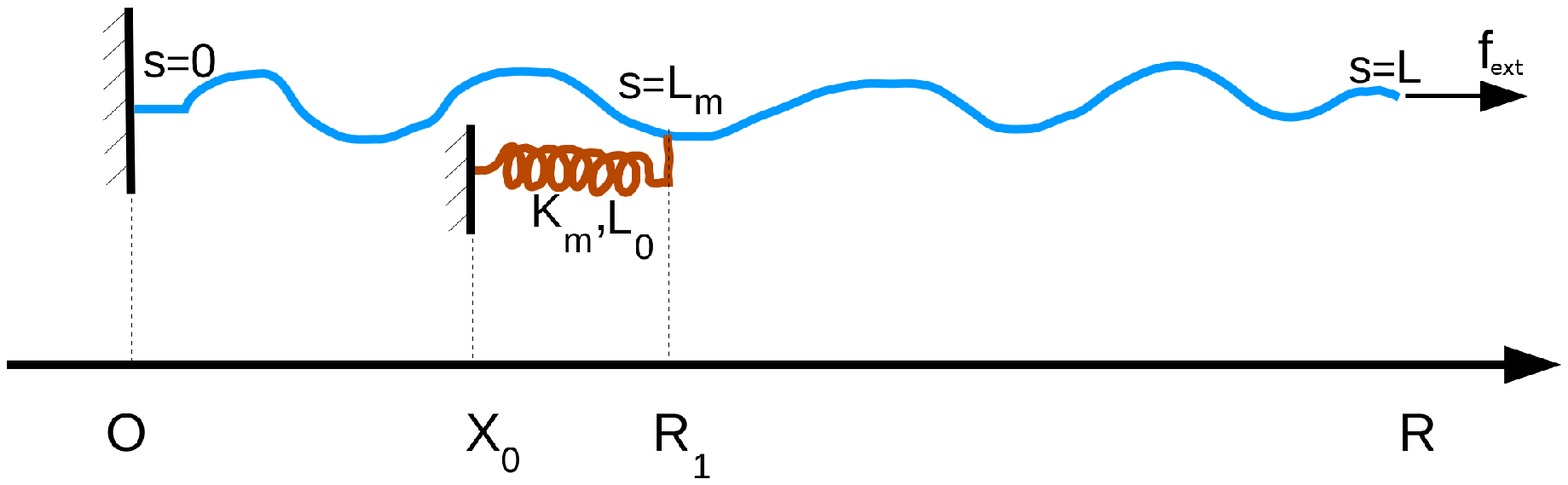}
\includegraphics[width=0.50\textwidth]{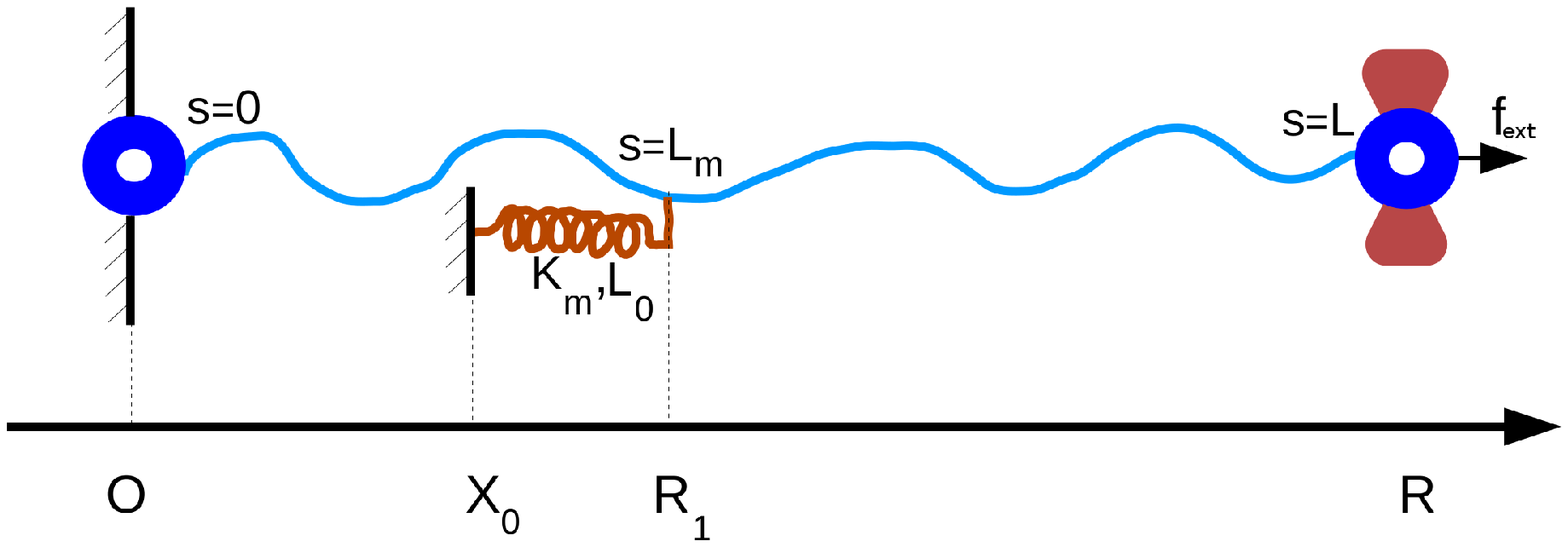}

}
\caption{Schematic presentation of a filament attached to a spring with clamped-free and hinged-hinged boundary conditions at the ends. One end of the spring is fixed to the substrate. The spring pulls one side of the filament and pushes the other side.
Therefore, the two sides of the filament are at different tension.}
\label{configurationofthesystem1}
\end{figure}

Here, we are interested in the effects of a motor whose one end is
grafted to a substrate while the other end is attached to the filament
at arclength $L_m$. In the simplest model, the motor is just a spring
of rest length $L_0$. and spring constant $k_m$
\begin{equation}
H_{spring}=\frac{k_m}{2}(R_1-X_0-L_0)^2
\end{equation}
 When the motor steps along the filament, its
effective spring is compressed or extended beyond the rest length, resulting in a
force, $f_m$ on the filament. Since the pulling force, $f_{ext}$, is fixed
the presence of the motor will result in a compression or extension of
the filament. In this paper, we compute the change in the
force-extension relation of the filament due to the attached motor.

In the weakly bending approximation, we can represent the total
Hamiltonian $H=H_{WB}+H_{spring}$ in the form of a filament with an
arclength dependent tension:
 \begin{eqnarray}
\label{Hamiltonian}
H=\sum_{i=1}^{2}\left[\int_{0}^{L}\left(\frac{\kappa}{2}\dot{a}^2_i(s)+\frac{f(s)}{2}a_i^2(s)\right) ds\right]
\end{eqnarray}
where $f(s)$ is a piecewise constant function
\begin{equation}
f(s)=
 \begin{cases}
f_{ext}+f_m & 0<s<L_m \\
f_{ext} & L_m<s<L
\end{cases}
\end{equation}
with $f_m=k_m(X_0-L_m+L_0)$.

Actually the assumtion of a harmonic spring for the motor is not
needed as long as we use the weakly bending approximation. Consider a
general interaction potential instead $V(R_1)$. In the weakly
bending approximation, we take
 \begin{equation}\label{R1hh}
R_1 -L_m=-\frac{1}{2}\sum_{i=1}^{2}\int_{o}^{L_m} a^2_i(s) ds
\end{equation}
to be small, and expand $V$ around $R_1=L_m$
\begin{eqnarray}
V(R_1)&=&V(L_m)+\frac{\partial V}{\partial R_1}(R_1-L_m)\nonumber\\
&=& V(L_m)+\frac{f_m}{2}\sum_{i=1}^{2}\int_{0}^{L_m}a_i^2(s) ds\nonumber
\end{eqnarray}
resulting in the same effective Hamiltonian (Eq.\ref{Hamiltonian}), but now for a general
interaction potential.

We want to compute the end-to-end distance of the filament $\langle R \rangle = \langle x(L)-x(0) \rangle$.
To that end, we first calculate % $\langle R_1 \rangle = \langle x(L_m)\rangle $
% \begin{equation}\label{R1hh}
%\left \langle R_1 \right \rangle=L_m-\frac{1}{2}\sum_{i=1}^{2}\int_{o}^{L_m}\langle a^2_i(s)\rangle ds
%\end{equation}
%and similarly  $\langle R_2 \rangle =\langle x(L)\rangle -\langle x(L_m)\rangle $
 \begin{equation}\label{R2hh}
\left \langle R_2 \right \rangle=(L-L_m)-\frac{1}{2}\sum_{i=1}^{2}\int_{L_m}^{L}\langle a^2_i(s)\rangle ds
\end{equation}
and similarly $\left \langle R_1 \right \rangle$,
where the thermal average $\langle ...\rangle$ is to be taken with the Hamiltonian
of Eq.(\ref{Hamiltonian}).

\section{Solution using Green function}

Using integration by parts for the first term in
Eq. (\ref{Hamiltonian}) we find:
 \begin{equation}
H_{WB}=\frac{1}{2}\sum_{i=1}^{2}\left[\int_{0}^{L}a_i(s)O(s)a_i(s) ds+B_i\right]
\end{equation}
where $B_i=\frac{\kappa}{2}a_i(s)\dot{a}_i(s)\mid_0^L$ depends on the
boundary conditions and is a constant and
$O(s)=-\kappa\frac{d^2}{ds^2}+f(s)$ is a differential operator. 
The corresponding Green function  obeys the differential equation
\begin{equation}
\beta (-\kappa\frac{d^2}{ds^2}+f(s))G_{hh}(s,s')=\delta(s-s')
\end{equation}
For a piecewise constant force $f(s)$ we can solve for the Green
function in the two regions with constant force and then match the
solutions at $s=L_m$ (see appendix). For the explicit calculation, we
have to specify boundary conditions at both ends of the filament. We
consider two cases: the clamped-free filament, shown in the upper part
and the hinged-hinged filament, shown in the lower part of
Fig.\ref{configurationofthesystem1}.

\subsection{Clamped-free filament}

We require $\dot{a}_i(L)=0$ at the free end and ${a}_i(0)=0$ at the
clamped end. For the Green function this implies
\begin{equation}
G_{cf}(s,s')\mid _{s=0}=0\quad
\frac{\partial }{\partial s}G_{cf}(s,s')\mid _{s=L}=0
\end{equation}
The correlation function of the transverse components of the tangent
vector can be obtained from the Green function as follows \cite{hori2007stretching,peskin1995introduction}:
 \begin{equation}
\langle a_i(s) a_i(s')\rangle =G_{cf}(s,s')
 \end{equation}

If no motor is attached, the force-extension relation reads \cite{PhysRevE.74.041803}:
\begin{equation}\label{Marko-Siggia-cf}
\left \langle R \right \rangle_{WLC}=\left \langle R_1+R_2 \right \rangle=L-\frac{L^2}{2l_p}(\frac{\tanh(\tilde{f}_{ext})}{\tilde{f}_{ext}})
\end{equation}
The characteristic energy scale of the WLC is given by
$\kappa/L$. Hence we have rescaled the externally applied pulling
force with the bending force of the wormlike chain and introduced
$\tilde{f}_{ext}=f_{ext} L^2/\kappa$. We get a linear relation for small forces
% $\left \langle R \right \rangle=L-\frac{L^2}{2lp}(1-\frac{1}{3}\sigma^2 L^2)$ 
%or:
\begin{equation}\label{linear-cf}
 f_{ext}=k_{\parallel}^{cf}(L)(L_r-\left \langle R \right \rangle)
\end{equation}
with rest length $L_r=L-L^2/2l_p$ and stiffness
$k_{\parallel}^{cf}(L)=\frac{6l_p^2}{\beta L^4}$ \cite{PhysRevE.74.041803}.

\subsection{Hinged-hinged filament}

In this case we require $\dot{a}_i(L)=0$ and $\dot{a}_i(0)=0$ 
implying for the Green function 
\begin{equation}
\frac{\partial }{\partial s}G_{hh}(s,s')\mid _{s=0}=0\quad
\frac{\partial }{\partial s}G_{hh}(s,s')\mid _{s=L}=0.
\end{equation}
Fo a compressive external force $f_{ext}$, the filament is free to rotate at the
grafted end. This can be prevented by requiring that the pulling point
has to have the same height as the grafting point: $\int_{0}^{L}a_i(s)=0$.
The correlation function of the components of the tangent vector
is then given by\cite{hori2007stretching,peskin1995introduction}:
 \begin{eqnarray}\label{correlation-hh}
\left<a_i(s)a_i(s')\right> &=-\frac{\int_{o}^{L}G_{hh}(s,s_1)ds_1\int_{o}^{L}G_{hh}(s',s_2)ds_2}{\int_{0}^{L}\int_{0}^{L}G_{hh}(s_1,s_2)ds_1ds_2}\nonumber\\
&+G_{hh}(s,s')
\end{eqnarray}

If no motor is attached to the filament, the force-extension is
explicitly given by
\begin{equation}\label{Marko-Siggia-hh}
\left \langle R \right \rangle_{WLC}=L-\frac{L^2}{2l_p}\left(\frac{\coth(\tilde{f}_{ext})}{\tilde{f}_{ext}}-\frac{1}{\tilde{f}_{ext}^2}\right).
\end{equation}
For small forces, the filament behaves
like a spring
\begin{equation}\label{linear-hh}
 f_{ext}=k_{\parallel}^{hh}(L)(L_r-\left \langle R \right \rangle)
\end{equation}
with rest length $L_r=L-L^2/(6l_p)$ and a length dependent stiffness
$k_{\parallel}^{hh}(L)=\frac{90(l_p)^2}{\beta L^4}$

\section{Results}

\begin{figure}[h]
\centering{
\includegraphics[width=0.45\textwidth]{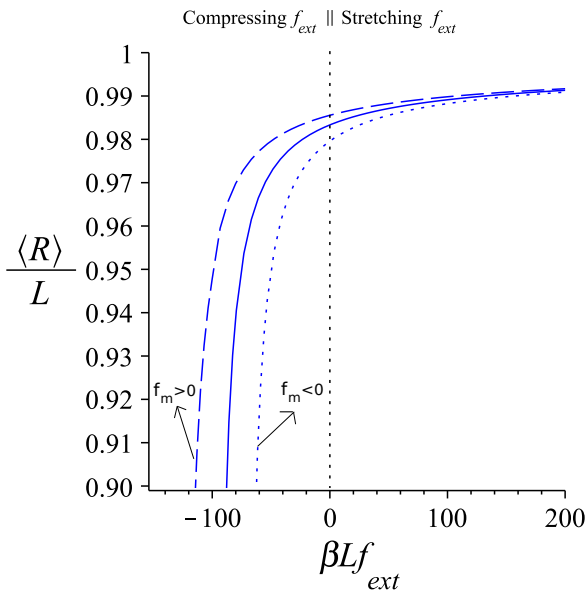}
}
\caption{Extension of the filament, $\langle R \rangle$, as a function
  of the external force $f_{ext}$ which can be compressive or extensile;
full line: no motor attached; dashed line $\beta L f_m=+50$;
dotted line: $\beta L f_m=-50$;
(parameters: $L=1$, $l_p=10$, $L/L_m=2$; hinged-hinged filament)}
\label{force-extension1}
 \end{figure}

\begin{figure}[h]
\centering{
\includegraphics[width=0.45\textwidth]{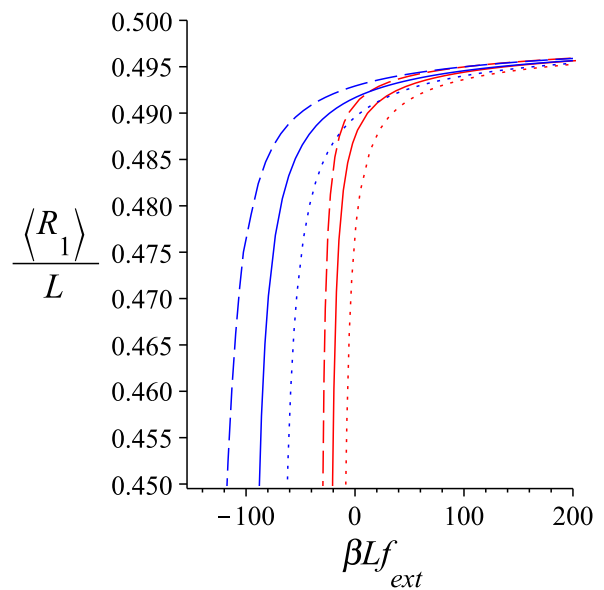}
\includegraphics[width=0.45\textwidth]{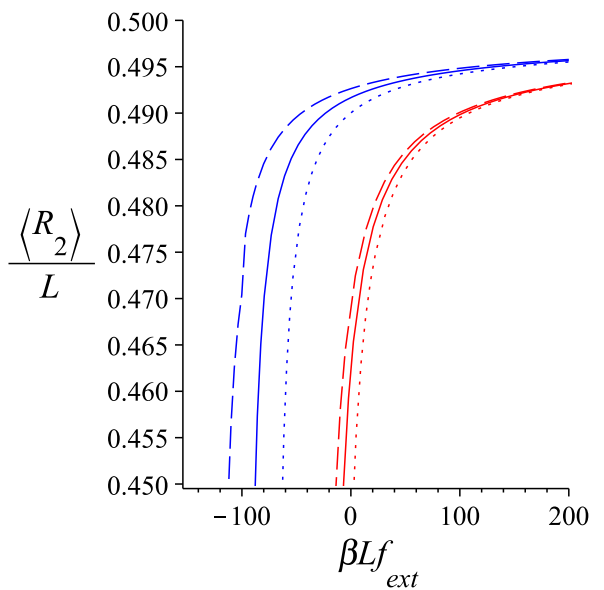}
}
\caption{Extension of the two separate parts of the filament,
  $ \langle R_1 \rangle$ and $ \langle R_2 \rangle$, as a function of
  the external force, $f_{ext}$, for several values of the motor
  force; red lines: clamped-free filament; blue lines: hinged-hinged
  filament. (full lines: no motor attached; dashed lines
  $\beta L f_m=+50$; dotted lines: $\beta L f_m=-50$; parameters as in
  Fig.\ref{force-extension1})}
\label{force-extension2}
 \end{figure}
 
\begin{figure}[h]
\centering{
\includegraphics[width=0.45\textwidth]{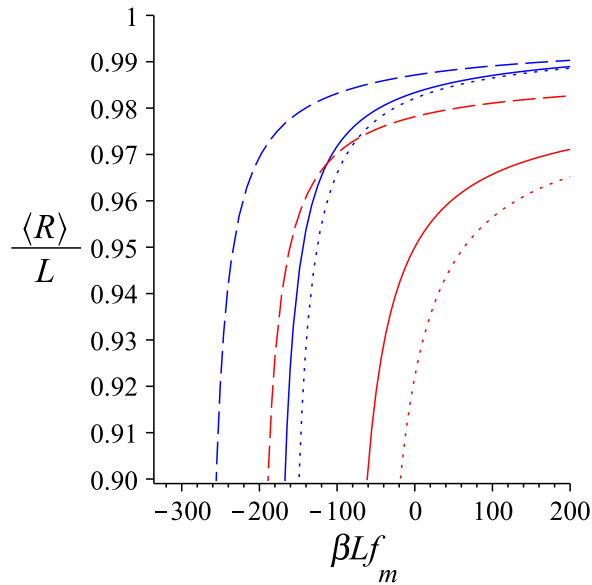}
}
\caption{Extension of the filament, $\langle R \rangle$, as a function
  of the motor force $f_m$ for several values of the external force;
  red lines: clamped-free filament, blue lines: hinged-hinged
  filament. (full lines: no motor attached; dashed lines
  $\beta L f_{ext}=+50$; dotted lines: $\beta L f_{ext}=-10$; parameters as in
  Fig.\ref{force-extension1})}
\label{force-extension3}
 \end{figure}

\begin{figure}[h]
\centering{
\includegraphics[width=0.45\textwidth]{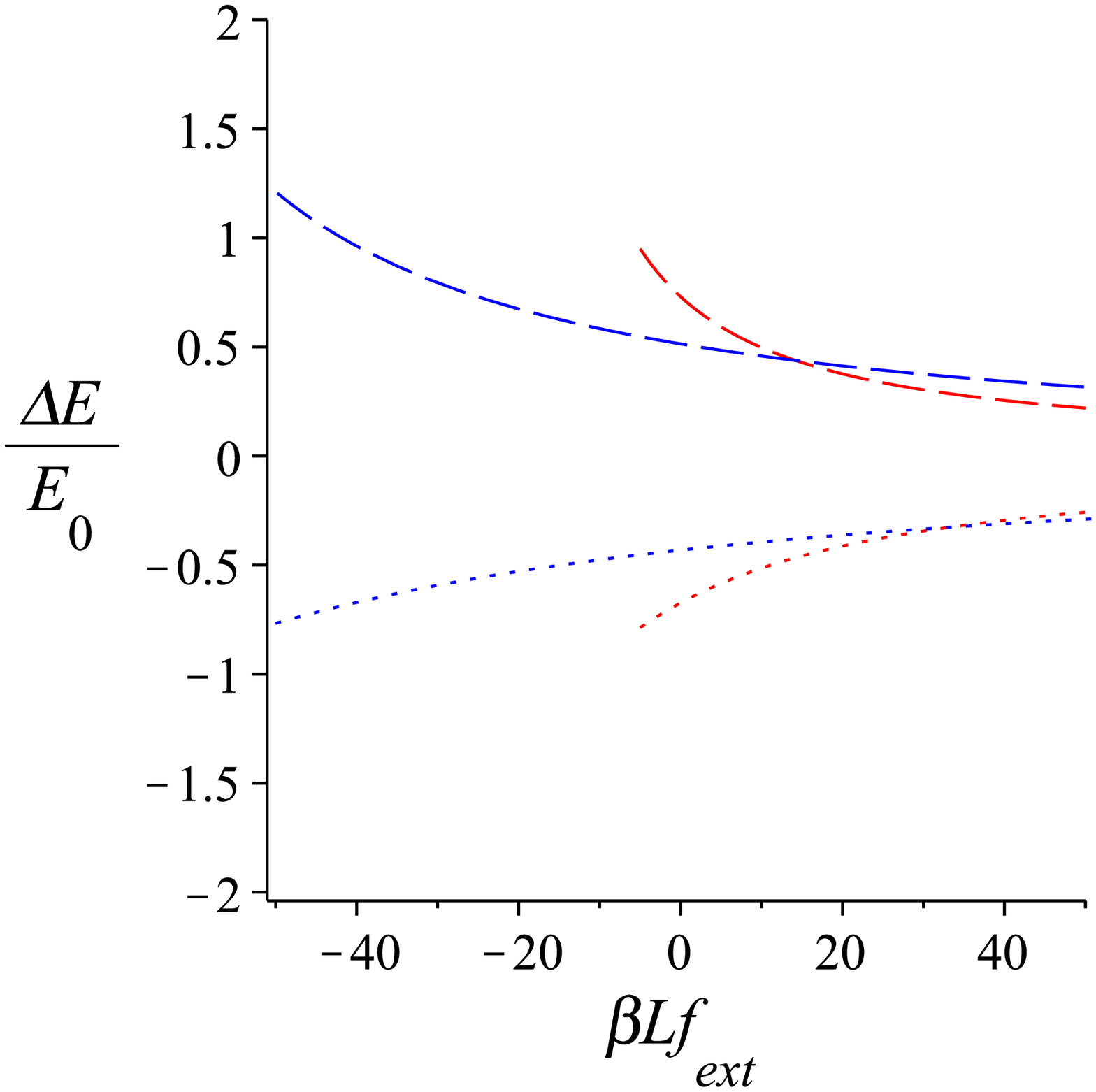}
}
\caption{ Change in the differential stiffness relative to the case
  without a spring as a function of the external force; red lines:
  clamped-free filament, blue lines: hinged-hinged filamen; (dashed
  lines $\beta L f_m=+50$; dotted lines: $\beta L f_m=-50$; parameters
  as in Fig.\ref{force-extension1}).}
\label{diff_stiffness}
 \end{figure}

 The explicit analytic solution for the Green function is given in the
 appendix. As a result we obtain analytic, albeit lengthy expressions
 for the force extension
 $\langle R \rangle = \langle R \rangle (f_{ext})$. To better
 understand these results, we plot the force-extension relations for
 hinged-hinged boundary conditions in Fig.~\ref{force-extension1}.
% and  \ref{force-extension2}.
%. Fig.~\ref{force-extension1} shows the dependence of the
% extension of the two pieces of the filament on the external pulling
% force, $f_{ext}$.
 The effect of the motor force is more pronounced in the compressional
 regime, $f_{ext}<0$, because the filament is softer in response to
 compressions as compared to extensions, $f_{ext}>0$. The motor force
 can partially compensate the compression of the fiber by the external
 force, if its sign is opposite, i.e. it is pulling on the left part
 of the segment. Obviously, the left segment is then extended as
 compared to the case without motor (see Fig.\ref{force-extension2}, upper part), but also the right segment is
 extended (see Fig.\ref{force-extension2}, lower part), even though the tension in the right part does not depend
 on $f_m$. The reason for this extension is the stronger alignment of
 the left end of the right part of the filament by the motor.  The
 overall effect of a positive motor force is to substantially stiffen
 the filament in the compressional regime.  The effects of course
 increase with increasing motor force. If the motor force is
 compressional the extension of the filament is correspondingly
 reduced as compared to the case without motor force.

 In Fig.~\ref{force-extension3}, we show the dependence of the
 filaments extension on the motor force, $f_m$, for several values of
 external pulling force, $f_{ext}$. If no external force is applied
 $\langle R \rangle = \langle R \rangle (f_m)$ looks qualitatively
 similar to $\langle R \rangle = \langle R \rangle (f_{ext})$. The
 filament is most sensitive to the external pulling force in the range
 where the motor tends to compress the filament.%  Again we observe a
 %stiffening of the filament which is more pronounced for smaller
 %external pulling force $f_{ext}$.

 The effects of the motor are seen best in the differential tensile
 stiffness of the filament, which can be computed from the
 force-extension relation according to
\begin{equation}
E^{-1}=\frac{\partial \langle R \rangle }{\partial f_{ext}}
\end{equation}
In Fig.~\ref{diff_stiffness} we show the relative change in the
differential stiffness of the filament caused by the spring. There is
significant enhancement in stiffness when the spring force is
extensile (dashed line), because the effective tension of the filament is
increased. The stiffness is weakened for a compressive motor force
(full lines). In both cases do we observe stronger effects in the
regime where the external force is compressive, implying that a
filament under compression is strongly sensitive to a motor which is
either pushing or pulling.

 \subsection{Limit of small motor force $f_m$}
\begin{figure}[h]
\centering{
\includegraphics[width=0.45\textwidth]{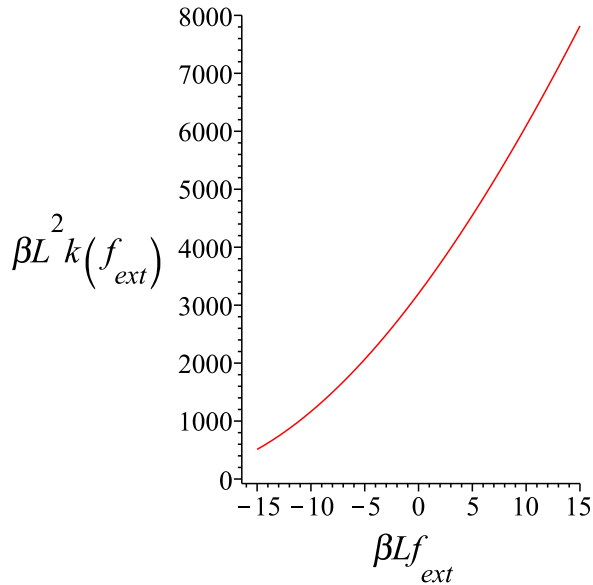}
}
\caption{ Effective stiffness, $k(f_{ext})$, of the filament with respect to the
  motor force as a function of external
  force $f_{ext}$.  The range of the external force is chosen in a way
  that the filament is well approximated as weakly bending
  ($\frac{R}{L}>0.9$); parameters as in Fig.\ref{force-extension1}).}
\label{kconst}
 \end{figure}

\begin{figure}[h]
\centering{
\includegraphics[width=0.45\textwidth]{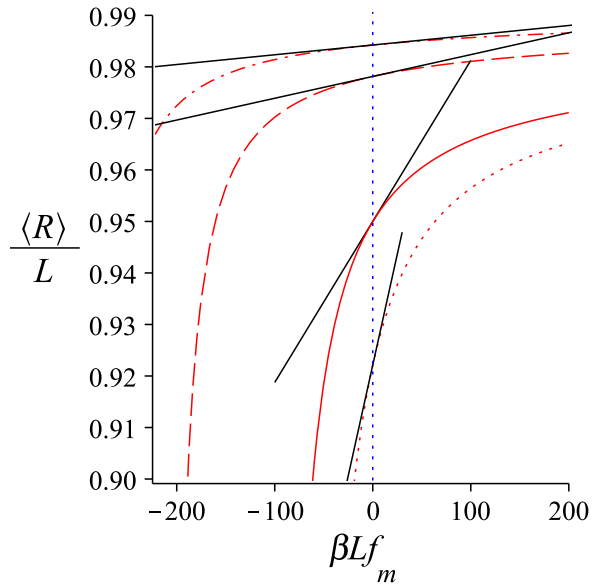}
}
\caption{Comparison of the complete solution for the end to end
  distance $\langle R \rangle/L$ and the one linearized around
  $f_m=0$;  full lines: $\beta L f_{ext}=0$; dashed lines
  $\beta L f_{ext}=50$; dashed-dotted lines: $\beta L f_{ext}=100$;
  dotted line: $\beta L f_{ext}=-10$ (parameters as in
  Fig.\ref{force-extension1}). }
\label{LinearG}
 \end{figure}

It is instructive to  consider the limit of small motor force
 $f_m\ll \min\{f_{ext},\frac{\kappa}{L^2}\}$. % where
 %$\tilde{L}\in \left\{L,L_m,(L-L_m)\right\}$.
For the clamped-free case, we find
\begin{equation}
\label{small_fm}
\left \langle R \right \rangle-\left \langle R \right \rangle_{f_m=0}
=\frac{f_m}{k(f_{ext})}.
\end{equation}
In this limit, the motor-filament system can be represented as a linear elastic element with an effective force constant 
%$k=k(\frac{f}{\kappa},l_p,L,L_m)$ 
$k$ that depends on the external pulling force and the point of
attachment of the motor. The explicit expression for the force
constant is given in the Appendix, and a similar expression can be
calculated for the hinged-hinged case. In Fig.~\ref{kconst}, we show
the force constant $k$ of the motor as a function of the external
force $f_{ext}$. The force constant decreases as we compress the
filament and it increases as we increase the stretching force.  This
change of the motor force constant $k$ is an essential feature of the
elasticity of the semiflexible filament which is missing in studies
using linear elasticity for the filament. In Fig.~\ref{LinearG}, we
compare the exact force-extension relation to the linearised one.  As
can be seen in the figure, the linear approximation works better for
higher values of the external force $f_{ext}$.

In the limit of large external forces, $ f_{ext}\gg  \max\{f_{m},\frac{\kappa}{{L_m}^2}\}$, and large filament length, $min\{L,L_m\}\gg f_{ext}/(k_BT)$, Eq. \ref{small_fm} reduces to the following
relation, irrespective of boundary conditions:
\begin{equation}
\label{large_f_thermodyn}
\frac{\left \langle R \right \rangle}{L}=1-\frac{1}{2}\frac{k_B T}{\sqrt{\kappa f_{ext}}}+\frac{L_m}{4L} \frac{k_B T f_m}{\sqrt{\kappa} {f_{ext}}^{\frac{3}{2}}}.
\end{equation}
Notice that this equation holds in the thermodynamic limit and it is
scale invariant: if we multiply all lengths
($\left \langle R \right \rangle$, $L_m$, $L$) by the same factor, it
does not change. The effect of the motor force is subdominant, as it
scales with $\sim 1/f_{ext}^{3/2}$ compared to $\sim 1/\sqrt{f_{ext}}$
for the Marko-Siggia case, but it is noteworthy that it persists in
the thermodynamic limit and is not just a finite-size effect.

In the limit of small external forces, $f_{ext}\ll \kappa/L^2$, we obtain a linear response to both the motor and the external force, which in the case of clamped-free boundary conditions reads:
\begin{equation}
\frac{\left \langle R \right \rangle_{cf}}{L}=1-\frac{L}{2l_p}+\frac{2L_m^3L-L_m^4}{6L^3l_p}\tilde{f}_m+\frac{L}{6l_p}\tilde{f}_{ext}\;,
\end{equation}

\subsection{Limit of large motor force $f_m$}
\begin{figure}[h]
\centering{
\includegraphics[width=0.45\textwidth]{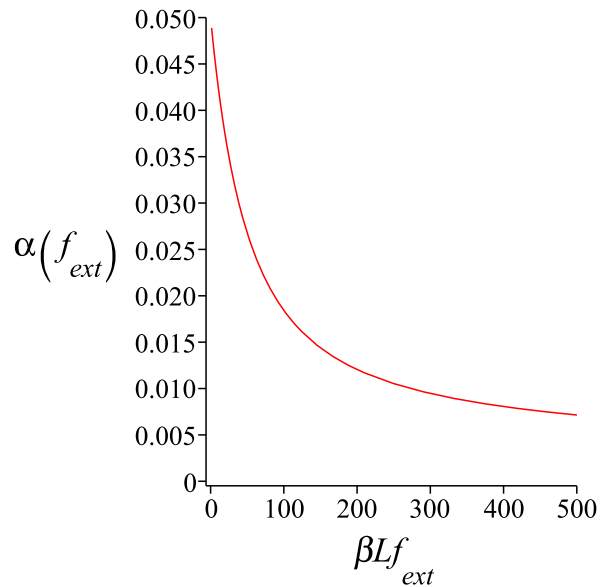}
}
\caption{Coefficient of the asymptotic expansion, $\alpha(f_{ext})$, as function of external force $f_{ext}$; (parameters as in
  Fig.\ref{force-extension1}). }
\label{LargeG}
 \end{figure}

In the limit of large motor force,
$f_m\gg \max\{f_{ext},\frac{\kappa}{{L_m}^2}\}$, we expect the left
part of the filament to display the asymptotic (Marko-Siggia) force
extension for large $f_m$ and indeed it does:
%$f_m\gg \max\{f_2,\frac{\kappa}{\tilde{L}^2}\}$, we have the following
%expression for the end to end distance of the first piece of the
%filament:
\begin{equation}
\left \langle R_1 \right \rangle_{cf}=L_m-\frac{L_m}{2l_p}\sqrt{\frac{\kappa}{f_m}}
\end{equation}
However the extension of the right part of the filament is not just
given by the expression for a wormlike chain under tension $f_{ext}$ but shows a correction of
${\cal O}(\frac{1}{\sqrt{f_m}})$ 
\begin{equation}
\left \langle R_2 \right \rangle_{cf}=\left \langle R_2 \right \rangle_{WLC}
-\alpha(f_{ext})\sqrt{\frac{\kappa}{f_m}}.
\end{equation}
The strength of the effect depends on the external pulling force (see
Fig.~\ref{LargeG}) and is strongest for weak pulling force. Explicit forms of $\left \langle R_2 \right \rangle_{WLC}$ and $\alpha(f_{ext})$ are given in the Appendix.

In the limit of small external force, $f_{ext}\ll \kappa/L^2 \ll f_m$ and $\kappa/L_m^2 \ll f_m$,
we obtain a linear force-extension relation:
\begin{eqnarray}
\frac{\left \langle R \right \rangle_{cf}}{L} &&=1-\frac{L}{2l_p}+\frac{L_m(6L-3L_m)}{6l_pL}+\frac{(L-L_m)^4}{6L^3l_p}\tilde{f}_{ext} \nonumber\\
&&-\frac{(6L-3L_m)L_m}{6l_pL}\frac{1}{\tilde{f}_m^\frac{1}{2}}\;.
\end{eqnarray}

\subsection{Limit of large force $f_{ext}$ and $f_m=-f_{ext}+\epsilon$}

Next, we consider the limit of a large external pulling force, which
is almost compensated by the motor in the left part of the
filament. In other words, we put $f_m=-f_{ext}+\epsilon$ and consider
the case with $f_{ext}\gg \frac{\kappa}{{L_m}^2}$ and
$\epsilon\ll \frac{\kappa}{{L}^2}$. In this limit, the right part of
the filament is asymptotically extended
\begin{equation}
\left \langle R_2 \right \rangle_{cf}=L-L_m-(\frac{L-L_m}{2l_p})
(\frac{\kappa}{f_{ext}})^{1/2}.
\end{equation}
The total extension in this limit is given by
\begin{equation}
\left \langle R \right \rangle_{cf}-L=\frac{L_m^2}{6l_p}-(\frac{L}{6l_p})\frac{(3L-L_m)}{\tilde{f}_{ext}^{\frac{1}{2}}}
+(\frac{L_m^4}{90l_pL^2})\tilde{\epsilon}.
\end{equation}
The first term accounts for the reduction in length due to thermal
fluctuations in the left part only and hence $\propto L_m^2$. The
tension in the left part is just $\epsilon\ll 1$ which accounts for
the last term. However the pulling force $f_{ext}$ affects the orientation
of the tangent at $L_m$ and hence also the extension of the left part
of the filament so the dominant term for
strong pulling force is not just determined by the right part of the
filament.

\section{relation to single-motor experiments}

Our results can be tested experimentally using optical tweezers. Beads
attached to the two ends of the biomolecule which acts as track (e.g.,
F-actin) are trapped with optical tweezers. The motor (e.g., myosin-V)
is attached with one end to a fixed bead and with the other end to the
filament. Since the two end beads are free to rotate, this arrangement
corresponds to the case of hinged-hinged boundary conditions. This
experimental set-up has already been used in many single-molecule
mechanical transducers \cite{Veigel_Schmidt_NR2011}. The "three-bead"
technique was pioneered by Finer {\it et al.} \cite{Finer_1994}. The
main idea is to measure the variance of the end-beads' position which
is related to the stiffness of the actomyosin cross-bridge complex
within linear elasticity \cite{Veigel_Molloy_2001,Veigel_1998}. Conformational
changes in the motor induce changes in the effective stiffness of the
bridge which is measured experimentally. In our model, the
conformational change in myosin changes the motor force $f_m$. This
can be due to a change in the position of attachment of the myosin
head on the actin filament, $L_m$, or a change in the effective spring
constant, $k_m$, or both.

Our results have been obtained in the fixed force ensemble, where the tension on the filament is determined sharply and this results in a fluctuating extension, whose average we have calculated. The positions of the motor bead, which determines $X_0$ in our model, and of the left filament end are held fixed. This can be done by using a very stiff optical trap. As shown by Gerland {\it et al.} in \cite{Gerland_Bundschuh_Hwa}, a polymer held between two optical traps  is represented by the mixed ensemble, where both the tension and the extension fluctuate. This mixed ensemble interpolates between the fixed extension ensemble (corresponding to the limit of very stiff traps) and fixed force ensemble (corresponding to the limit of very soft traps). Therefore, our general results for hinged-hinged boundary conditions can be tested with a set up involving a very soft optical trap for the right end of the filament. We should point out, that the force-extension relation  given by Eq. \ref{large_f_thermodyn} holds in the thermodynamic limit and as such is ensemble independent (fixed force or fixed extension). In addition, the linear response results are ensemble independent.

In real systems, the spring will act not only in the longitudinal
direction but also in the transverse direction. The effect of a
transverse spring of zero rest length in the force-extension relation
of a weakly bending wormlike chain has been calculated in
\cite{PB_SU_AZ_NJP}. For a spring of finite rest length which is
almost parallel to the longitudinal direction of the filament, the
transverse effect is of higher order and can be neglected. For a
spring of zero rest length but with $f_m\neq0$, we can simply add the
following contribution to the right-hand side of
Eq. \ref{large_f_thermodyn}:
\begin{eqnarray}
\label{transverse_spring}
\frac{\Delta \langle R \rangle}{L}=\frac{k_m k_BT}{4 f_{ext}^2}\Big(1+\frac{k_m L}{2f_{ext}}\Big)^{-1},
\end{eqnarray}
which holds in the strong stretching limit, $f\gg \kappa/L_m^2$. For a soft spring, $k_m L \ll f_{ext}$, this contribution falls off as $\sim f_{ext}^{-2}$, which is subdominant to the longitudinal contribution which falls off as $\sim f_{ext}^{-3/2}$.

\section{Conclusion-Outlook}

We have analysed the force-extension of a wormlike chain whose one end
is fixed, while the other end is pulled or pushed by an external
force. In addition, the filament is attached to a spring which may
represent a cross-link or a motor arrested at its stall force. % For
%clamped-free boundary conditions, the effects of the spring are
%strongest in the confined region of the filament between fixed end and
%spring which is softer compared to the free part. Also,
Irrespective
of boundary conditions, the effects of the spring are stronger in the
compressive regime as compared to the stretching regime. Depending on the
relative sign of the pulling force and the spring force, the latter
can substantially stiffen or weaken the filament. When the motor force
is small, its effects can be represented by an effective spring
constant which strongly depends on the prestress of the fiber, {\it
  i.e.} the external force. When the motor force is large, it gives
rise to the same $1/\sqrt{f_{ext}}$ dependence which is well known
from the work of Marko and Siggia \cite{Marko_Siggia}.

The dependence of the force extension curve of the filament on motor
force allows to deduce the latter from measurements of the force-extension relation. In fact the so-called three bead geometry has already been
used to determine the stiffness of the actomyosin
cross-bridge~\cite{Veigel_1998}.

An interesting direction for future work is the study of two or more
parallel-aligned filaments with non-local spring-like cross-linkers in
the direction of alignment. The case of local cross-links has been
investigated in \cite{Alice1,Alice2,PB_SU_AZ_NJP}. A simplified model
of two filaments with a non-local spring has been studied in
\cite{Liverpool_Marchetti_EPL2009}.

\section{Acknowledgement}
We thank the DFG for financial support through SFB 937, project A1.

\section{Appendix}
\subsection*{A.1. Filament with hinged-hinged boundary conditions at the two tips}
\begin{widetext}
The force is a piecewise
constant function with two pieces. As result of this fact, the Green function is a piecewise function with six pieces:

\begin{equation}\label{Green1}
G(s,s')=
\begin{cases} 
 G_{1}^{-}(s,s') & 0<s<s' \leq L_m<L \\
 G_{1}^{+}(s,s') & 0<s'<s \leq L_m<L \\
                                                               \\
 G_{2}^{-}(s,s') & 0<L_m \leq s<s'<L \\
 G_{2}^{+}(s,s') & 0<L_m \leq s'<s<L \\                                                              
                                                               \\
 G_{3}^{-}(s,s') & 0<s<L_m<s'<L \\
 G_{3}^{+}(s,s') & 0<s'<L_m<s<L \\                                                            
\end{cases}
\end{equation}
The assumption that the derivative of the tangent vector is zero at the end tips (hinged-hinged condition) leads to 
the vanishing of the derivative of the Green function.
 Considering the boundary conditions, the solution for the aforementioned equation must have  the following form:

\begin{equation}\label{Green2}
G_{hh}(s,s')=
\begin{cases} 
G_{1}^{-}(s,s')=N_{11}(s')\cosh(s\sigma_1) \\
G_{1}^{+}(s,s')=N_{12}(s')(\cosh(s\sigma_1)+A_1\sinh(s\sigma_1)) \\
                                                               \\
G_{2}^{-}(s,s')=N_{21}(s')(\cosh(s\sigma_2)+A_2\sinh(s\sigma_2)) \\
G_{2}^{+}(s,s')=N_{22}(s')\cosh((L-s)\sigma_2)  \\                                                              
                                                               \\
G_{3}^{-}(s,s')=N_{31}(s')\cosh(s\sigma_1)  \\
G_{3}^{+}(s,s')=N_{32}(s')\cosh((L-s)\sigma_2) \;, \\                                                            
\end{cases}
\end{equation}
where $\sigma(s)= \sqrt{\frac{f(s)}{\kappa}}=\begin{cases} \sigma_1=\sqrt{\frac{f_1}{\kappa}} & 0<s<L_m \\\\ \sigma_2=\sqrt{\frac{f_2}{\kappa}} & L_m<s<L \end{cases}$ and $L_m$ is 
the position of molecular motor in terms of the contour length. 
Moreover, constants appearing in eq.~\ref{Green2} are obtained from the following conditions:

\begin{equation}\label{Conditions}
\begin{cases} 

 \frac{\partial G_{1}^{+}(s,s')}{\partial s}\mid _{s=s'}-\frac{\partial G_{1}^{-}(s,s')}{\partial s}\mid _{s=s'}=-\frac{1}{\beta\kappa} &(1)\\
 G_{1}^{-}(s',s')=G_{1}^{+}(s',s')&(2)\\\\
  \frac{\partial G_{2}^{+}(s,s')}{\partial s}\mid _{s=s'}-\frac{\partial G_{2}^{-}(s,s')}{\partial s}\mid _{s=s'}=-\frac{1}{\beta\kappa}&(3)\\
 G_{2}^{-}(s',s')=G_{2}^{+}(s',s')&(4)\\\\
  G_{3}^{-}(L_m,s')=G_{2}^{-}(L_m,s')&(5)\\
 G_{3}^{+}(L_m,s')=G_{1}^{+}(L_m,s')&(6)\\\\
 \frac{\partial G_{3}^{-}(s,s')}{\partial s}\mid _{s=L_m} = \frac{\partial G_{2}^{-}(s,s')}{\partial s}\mid _{s=L_m}&(7)\\
 \frac{\partial G_{3}^{+}(s,s')}{\partial s}\mid _{s=L_m} = \frac{\partial G_{1}^{+}(s,s')}{\partial s}\mid _{s=L_m}&(8)\;.
 \end{cases}
\end{equation}
These conditions coming in Eq.~\ref{Conditions}, except number (7) and number (8), gives:
\begin{equation}
\begin{cases} 
 N_{11}(s')=-\frac{\cosh(\sigma_1 s')+A_1\sinh(\sigma_1 s')}{\sigma_1 l_p A_1}\\\\
 N_{12}(s')=-\frac{\cosh(\sigma_1 s')}{\sigma_1 l_p A_1}\\\\
 N_{21}(s')=\frac{\cosh(\sigma_2(L-s'))}{\sigma_2 l_p(\sinh(\sigma_2 L)+A_2\cosh(\sigma_2 L))}\\\\
 N_{22}(s')=\frac{\cosh(\sigma_2 s')+A_2\sinh(\sigma_2 s')}{\sigma_2 l_p(\sinh(\sigma_2 L)+A_2\cosh(\sigma_2 L))}\\\\
 N_{31}(s')=\frac{\cosh(\sigma_2(L-s')(\cosh(\sigma_2 L_m)+A_2\sinh(\sigma_2L_m)))}{\sigma_2 l_p \cosh(\sigma_1L_m)(\sinh(\sigma_2L)+A_2\cosh(\sigma_2L))}\\\\
 N_{32}(s')=-\frac{\cosh(\sigma_1 s')(\cosh(\sigma_1L_m)+A_1\sinh(\sigma_1 L_m))}{\sigma_1 l_p A_1\cosh(\sigma_2(L-L_m))}\;,
 \end{cases}
\end{equation} 
The conditions number (7) and number (8) of eq.~\ref{Conditions} give:

\begin{equation}
\begin{cases} 
 A_1=-\frac{\sigma_2\sinh(\sigma_2(L-L_m))\cosh(\sigma_1L_m)+\sigma_1\sinh(\sigma_1L_m)\cosh(\sigma_2(L-L_m))}
 {\sigma_2\sinh(\sigma_2(L-L_m))\sinh(\sigma_1L_m)+\sigma_1\cosh(\sigma_1L_m)\cosh(\sigma_2(L-L_m))}\\\\
 A_2=-\frac{\sigma_1\sinh(\sigma_1L_m)\cosh(\sigma_2L_m)-\sigma_2\sinh(\sigma_2L_m)\cosh(\sigma_1L_m)}
 {\sigma_1\sinh(\sigma_1L_m)\sinh(\sigma_2L_m)-\sigma_2\cosh(\sigma_2L_m)\cosh(\sigma_1L_m)}

 \end{cases}
\end{equation}

Concerning the boundary condition, the correlation function of the transverse components of tangent vector is written as follows:
 \begin{equation}\label{Corelhh}
\left<a_i(s) a_i(s')\right>_{hh}=\lim_{J_i\rightarrow 0}\frac{\delta^2 \ln\left(Z(J_i)\right)}{\delta J_i(s) \delta J_i(s')}\;,
 \end{equation}
 where $Z(J_i)=\int D\left \{ a_i(s) \right \}\delta(\int_{0}^{L}dsa_i(s))\exp(-\beta H_{WBA}+\int_{0}^{L}dsJ_i(s)a_i(s))$ is 
the generating functional with source term $J_i(s)$ and $\beta=\frac{1}{k_BT}$. The correlation function of the transverse components of the tangent vector
is obtained by the following expression \cite{hori2007stretching,peskin1995introduction}:
 \begin{eqnarray}\label{correlation-hh}
\left<a_i(s)a_i(s')\right> =G_{hh}(s,s')-\frac{\int_{o}^{L}G_{hh}(s,s_1)ds_1\int_{o}^{L}G_{hh}(s',s_2)ds_2}{\int_{0}^{L}\int_{0}^{L}G_{hh}(s_1,s_2)ds_1ds_2}
\end{eqnarray}
which implies
 \begin{eqnarray}\label{correlation-hh}
\left<{a_i}^2(s)\right> =G_{hh}(s,s)-\frac{\left(\int_{o}^{L}G_{hh}(s,s_1)ds_1\right)^2}{\int_{0}^{L}\int_{0}^{L}G_{hh}(s_1,s_2)ds_1ds_2}
\end{eqnarray}
\subsection*{A.2. Filament with clamped-free boundary conditions at the two end tips}
In the clamped-free case, we enforce the transverse components of the tangent vector of the filament at $s=0$ and their derivitive at 
$s=L$ to be zero. Similar to Appendix 1, we obtain the following expression 
for the Green function:
\begin{equation}\label{Green}
G_{cf}(s,s')=
\begin{cases} 
G_{1}^{-}(s,s') \equiv
\frac{\sinh(\sigma_1 s)(B_1\cosh(\sigma_1 s')+\sinh(\sigma_1 s'))}{l_p \sigma_1 B_1}  \\\\
G_{1}^{+}(s,s')\equiv
\frac{(B_1\cosh(\sigma_1 s)+\sinh(\sigma_1 s))\sinh(\sigma_1 s')}{l_p \sigma_1 B_1}  \\\\
\\
G_{2}^{-}(s,s')\equiv
\frac{\cosh(\sigma_2(L-s'))(B_2\cosh(\sigma_2 s)+\sinh(\sigma_2 s))}{l_p\sigma_2 B_2 \sinh(\sigma_2 L)+l_p\sigma_2\cosh(\sigma_2 L)}  \\\\
G_{2}^{+}(s,s')\equiv
\frac{\cosh(\sigma_2 s)(B_2\cosh(\sigma_2 s')+\sinh(\sigma_2 s'))}{l_p\sigma_2 B_2 \sinh(\sigma_2 L)+l_p\sigma_2\cosh(\sigma_2 L)} \\\\                                                              
\\
G_{3}^{-}(s,s')\equiv
\frac{\sinh(\sigma_1 s)\cosh(\sigma_2(L-s'))(B_2\cosh(\sigma_2L_m)+\sinh(\sigma_2 L_m))}{l_p\sigma_2 \sinh(\sigma_1 L_m)(B_2\sinh(\sigma_2 L)+\cosh(\sigma_2 L))} \\\\
G_{3}^{+}(s,s')\equiv
\frac{\cosh(\sigma_2(L-s))\sinh(\sigma_1 s')(B_1\cosh(\sigma_1 L_m)+\sinh(\sigma_1 L_m))}{l_p\sigma_1 B_1 \cosh(\sigma_2(L-L_m))}  \\ 
                                                       
\end{cases}
\end{equation}
where:
\begin{equation}
\begin{cases} 
 B_1=-\frac{\sigma_2 \sinh(\sigma_2(L-L_m))\sinh(\sigma_1 L_m)+\sigma_1\cosh(\sigma_2(L-L_m))\cosh(\sigma_1 L_m)}
 {\sigma_2\sinh(\sigma_2(L-L_m))\cosh(\sigma_1 L_m)+\sigma_1\cosh(\sigma_2(L-L_m))\sinh(\sigma_1 L_m)}\\\\
 B_2=\frac{\sigma_1 \cosh(\sigma_1 L_m)\sinh(\sigma_2 L_m)-\sigma_2\sinh(\sigma_1 L_m)\cosh(\sigma_2 L_m)}
 {\sigma_2 \sinh(\sigma_1 L_m)\sinh(\sigma_2 L_m)-\sigma_1\cosh(\sigma_1 L_m)\cosh(\sigma_2 L_m)}

 \end{cases}
\end{equation}
The correlation function  \cite{hori2007stretching,peskin1995introduction}
is obtained as follows:
 \begin{equation}\label{correlation-hf}
\left<a_i(s)a_i(s')\right>_{cf}=G_{cf}(s,s')
\end{equation}
which implies
\begin{equation}
 \left<a^2_i(s)\right>_{cf}=G_{cf}(s,s)
\end{equation}
\subsection*{A.3 ~Linear end to end distance in terms of $f_m$}
The end to end distance of the whole filament when it is in the clamped-free condition can be written as follows:
\begin{equation}
\left \langle R \right \rangle_{cf}=L-\frac{L^2 \tanh(\tilde{f}_{ext}^{\frac{1}{2}})}{2l_p \tilde{f}_{ext}}+\left(\frac{1}{k(f_{ext})}\right){f_m}
\end{equation}
The effective linear motor force constant k is:
\begin{eqnarray}
\frac{1}{k}&&=-\frac{1}{4\left(l_p\kappa(\frac{f_{ext}}{\kappa})^{\frac{5}{2}}(e^{4L\sqrt{\frac{f_{ext}}{\kappa}}}+2e^{2L\sqrt{\frac{f_{ext}}{\kappa}}}+1)\right)}\\ \nonumber
\times&& \left[Ae^{4L\sqrt{\frac{f_{ext}}{\kappa}}}+Be^{2(2L-L_m)\sqrt{\frac{f_{ext}}{\kappa}}}+Ce^{2L\sqrt{\frac{f_{ext}}{\kappa}}}+
+De^{2(L+L_m)\sqrt{\frac{f_{ext}}{\kappa}}}+Ee^{2(L-L_m)\sqrt{\frac{f_{ext}}{\kappa}}}+Fe^{2L_m\sqrt{\frac{f_{ext}}{\kappa}}}+G
\right]\;,
\end{eqnarray}
where $A=-\frac{f_{ext} L_m}{\kappa}+\sqrt{\frac{f_{ext}}{\kappa}}, B=-\left(\frac{f_{ext} L_m}{\kappa}+\sqrt{\frac{f_{ext}}{\kappa}}\right), C=\left(\frac{4 f_{ext} L L_m}{\kappa}+2\right)\sqrt{\frac{f_{ext}}{\kappa}}, 
D=-\left(\sqrt{\frac{f_{ext}}{\kappa}}-\frac{f_{ext} L_m}{\kappa}+\frac{f_{ext} L}{\kappa} \right), E=-\sqrt{\frac{f_{ext}}{\kappa}}-\frac{f_{ext} L_m}{\kappa}+\frac{f_{ext} L}{\kappa}, F=\frac{f_{ext} L_m}{\kappa}-\sqrt{\frac{f_{ext}}{\kappa}}, G=\frac{f_{ext} L_m}{\kappa}+\sqrt{\frac{f_{ext}}{\kappa}}$  
\subsection{A.4~Limit of large motor force $f_m$}
In the limit of large motor forces $f_m\gg Max\{f_{ext},\frac{\kappa}{{L}^2}\}$, we have the following expression for the end to end distance of the first piece of the filament:
\begin{equation}
\left \langle R_1 \right \rangle_{cf}=L_m-\frac{L_m}{2l_p}\sqrt{\frac{\kappa}{f_m}}
\end{equation}
and the end to end distance of the second piece of the filament:
\begin{eqnarray}
\left \langle R_2 \right \rangle_{cf}&=\left(L-L_m\right)-\frac{\left(L-L_m\right)}{2l_p}\frac{\tanh\left(\left(L-L_m\right)\sqrt{\frac{f_{ext}}{\kappa}}\right)}{\sqrt{\frac{f_{ext}}{\kappa}}}\\ \nonumber
&-\alpha(f_{ext})\sqrt{\frac{\kappa}{f_m}}\;,
\end{eqnarray}
where
\begin{eqnarray}
\alpha(f_{ext})&=\frac{e^{4L\sqrt{\frac{f_{ext}}{\kappa}}}-e^{4L_m\sqrt{\frac{f_{ext}}{\kappa}}}}{2l_p\sqrt{\frac{f_{ext}}{\kappa}}\left(e^{2L\sqrt{\frac{f_{ext}}{\kappa}}}+e^{2L_m\sqrt{\frac{f_{ext}}{\kappa}}}\right)^{2}}\\ \nonumber
&+\frac{2(L-L_m)e^{2(L+L_m)\sqrt{\frac{f_{ext}}{\kappa}}}}{l_p\left(e^{2L\sqrt{\frac{f_{ext}}{\kappa}}}+e^{2L_m\sqrt{\frac{f_{ext}}{\kappa}}}\right)^{2}}
\end{eqnarray}
\end{widetext}
\bibliographystyle{unsrt}
\bibliography{testbibnew}
\end{document}